\documentclass[prd,amsmath,amssymb,aps,twocolumn]{revtex4-1}

\usepackage{color}

\newcommand{\subE}{\textrm{\tiny{E}}}
\newcommand{\subS}{\textrm{\tiny{S}}}
  \newcommand{\be}{\begin{equation} }
 \newcommand{\ee}{\end{equation}}
    \newcommand{\bes}{\begin{equation*} }
 \newcommand{\ees}{\end{equation*}}
  \newcommand{\bea}{\begin{eqnarray} }
 \newcommand{\eea}{\end{eqnarray}}
    \newcommand{\beas}{\begin{eqnarray*} }
 \newcommand{\eeas}{\end{eqnarray*}}
   \newcommand{\ba}{\begin{align} }
 \newcommand{\ea}{\end{align} }
  \newcommand{\bas}{\begin{align*} }
   \newcommand{\eas}{\end{align*} }

\usepackage{graphicx}
\usepackage{multirow}
\usepackage[colorlinks=true,citecolor=blue]{hyperref}

\begin{document}

\title{A Mode-Sum Prescription for Vacuum Polarization in Odd Dimensions}

\author{Peter Taylor}
\email{peter.taylor@ucd.ie}
\affiliation{Center for Radiophysics and Space Research, Cornell University, Ithaca, NY 14853, USA\\
 and\\
School of Mathematical Sciences, University College Dublin, UCD, Belfield, Dublin 4, Ireland}
\author{Cormac Breen}
\email{cormac.breen@dit.ie}
\affiliation{School of Mathematical Sciences, Dublin Institute of Technology, Kevin Street, Dublin 8, Ireland}

\date{\today}
\begin{abstract}
We present a new mode-sum regularization prescription for computing the vacuum polarization of a scalar field in static spherically-symmetric black hole spacetimes in odd dimensions. This is the first general and systematic approach to regularized vacuum polarization in higher dimensions. Remarkably, the regularization parameters can be computed in closed form in arbitrary dimensions and for arbitrary metric function $f(r)$. In fact, we show that inspite the increasing severity and number of the divergences to be regularized, the method presented is mostly agnostic to the number of dimensions. Finally, as an explicit example of our method, we show plots for vacuum polarization in the Schwarzschild-Tangherlini spacetime for odd $d=5,...,11$. \end{abstract}
\maketitle

\section{Introduction}
Quantum gravity remains one of the most important outstanding problems in physics. Abesent a full theory, one must rely on approximations. One particularly important approximation is semi-classical gravity, which is the treatment of quantum fields interacting with a classical spacetime metric via the semi-classical Einstein equations
\begin{align}
	G_{ab}=8\pi \langle T_{ab}\rangle.
\end{align}
The source term in these equations is the expectation value of the stress-tensor of the quantum fields being considered.

Solving the semi-classical Einstein equations is notoriously difficult. The first major obstacle one encounters is that the source term is in fact divergent, being quadratic in an operator-valued distribution. A formal prescription to regularize the stress-energy tensor--the point-splitting scheme--dates back to DeWitt and Christensen \cite{dewitt1975quantum, ChristensenPointSplit}. Effectively, the prescription amounts to considering the stress-tensor evaluated at two nearby spacetime points and then subtracting a parametrix that encodes all the geometrical divergences in the coincident limit. Applying the point-splitting scheme in a way that is amenable to numerical evaluation still remained a challenge, the first work in this direction was the seminal work of Candelas and Howard \cite{CandelasHowardPhi2}. Notwithstanding the ingenuity of their method, this approach has some serious drawbacks, including its crucial dependence on WKB methods--which are problematic in the Lorentzian sector--and its lack of numerical efficiency. It has also proved difficult to generalize to spacetimes that are not highly symmetric. Nevertheless, the Candelas-Howard approach has remained more or less the standard prescription for several decades.

Departures from the Candelas-Howard method are sparse in the literature. We mention a couple of examples. Ottewill and Taylor \cite{OttewillTaylorCS} devised a regularization scheme on the Schwarzschild spacetime threaded by a cosmic string. The method involves generating a mode-sum expression for the Hadamard parametrix by a clever matching to the flat spacetime Green function. A more recent endeavour by Levi and Ori \cite{LeviOriMSP} has resulted in methods that seem to be applicable to more general spacetimes with relaxed symmetry assumptions, though results have only been reported for static, spherically symmetric cases. Here we present a new systematic scheme for higher dimensions. We restrict our attention to vacuum polarization for a scalar field on static, spherically-symmetric black hole spacetimes in arbitrary odd dimensions.  We present the even-dimensional case in a separate article \cite{TaylorBreenEvenD}; it is more complicated because of the presence of log terms in the singular two-point function. We note also that the methods developed in this series of papers should readily extend to the more technically challenging computation of the renormalised stress energy tensor.

While the renormalised vacuum polarization has been computed for a variety of black hole spacetimes in four dimensions, see for example \cite{Candelas:1980zt,CandelasHowardPhi2,Candelas:1985ip,Anderson:1989vg,DeBenedictis:1998be,Frolov:1982pi,breen:10,Winstanley:2007tf,OttewillTaylorCS,Flachi:2008sr}, there has been comparatively little work carried out in higher dimensional black hole spacetimes. Some general properties of the renormalised stress energy tensor in higher dimensions are derived in \cite{Morgan:2007hp} while Decanini and Folacci \cite{DecaniniFolacciHadamardRen} developed a formalism for its computation based on Hadamard renormalization. Another general formalism for computing the renormalised vacuum polarization in higher dimensional spacetimes based on DeWitt-Schwinger renormalization was developed in \cite{Thompson:2008bk}. Large mass approximations based on the DeWitt-Schwinger expansion have also been computed and used to study back-reaction effects in higher-dimensional black hole spacetimes \cite{MatyjasekHD1, MatyjasekHD2}. For exact renormalized quantities, Ref.~\cite{Fro:89} presents the renormalized vacuum polarization on the event horizon of a five-dimensional Schwarzschild-Tangherlini black hole. Very recently, Flachi \textit{et al.} \cite{Flachi:HD2016} have presented a numerical calculation of the vacuum polarization on the exterior of the five-dimensional Schwarzschild-Tangherlini spacetime, the calculation is based on the Candelas-Howard method which is cumbersome and inefficient to generalize to higher dimensions. As far as we are aware, the only other computations of the renormalised vacuum polarization in the exterior region of higher-dimensional black holes is in the context of braneworld models where one considers quantum effects on the four-dimensional brane of a higher-dimensional bulk spacetime (see, e.g., \cite{breen2015vacuum,shiraishi1994vacuum}).  

Our approach is in a sense the most natural and direct approach to the problem: a full multipole and Fourier decomposition of the Hadamard parametrix. For a judicious choice of separation variables in which to expand the Hadamard parametrix, the coefficients in this decomposition--which we call regularization parameters--can be computed once-and-for-all in arbitrary dimensions, providing an out-of-the-box solution for regularization in static, spherically-symmetric spacetimes. This results in a complete mode-by-mode sum for the regularized vacuum polarization. Moreover, the regularization parameters to any desired order--corresponding to any order in the Hadamard parametrix--can be computed. This is extremely useful since inclusion of higher order terms in the parametrix speeds the convergence of what is typically a very slowly convergent mode-sum. Here, we include all terms up $\textrm{O}(\epsilon)$ in our decomposition of the singular parametrix, where $\epsilon$ sclaes like the distance between two nearby points. So efficient is the resultant mode-sum, that its numerical evaluaton is completely straight-forward, requiring only a few tens of modes to attain accuracies of approximately ten decimal places. We demonstrate the utility and efficiency of the method by giving explicit plots for vacuum polarization in the Schwarzschild-Tangherlini spacetime for odd dimensions between $d=5,...,11$.

\section{The Euclidean Green Function}
We consider a quantum scalar field on a static, spherically symmetric black hole spacetime of the form
\begin{align}
	\label{eq:metric}
	ds^{2}=-f(r)dt^{2}+dr^{2}/f(r)+r^{2}d\Omega^{2}_{d-2},
\end{align}
where $d\Omega^{2}_{d-2}$ is the metric on $\mathbb{S}^{d-2}$. Assuming the field is in a Hartle-Hawking state, we can adopt Euclidean techniques to simplify the problem. In particular, performing a Wick rotation $t\to-i\,\tau$ results in the Euclidean metric
\begin{align}
	ds^{2}=f(r)d\tau^{2}+dr^{2}/f(r)+r^{2}d\Omega_{d-2}^{2}.
\end{align}
It can be shown that this metric would possess a conical singularity unless we enforce the periodicity $\tau=\tau+2\pi/\kappa$ where $\kappa$ is the surface gravity. This discretizes the frequency spectrum of the field modes which now satisfy an elliptic wave equation
\begin{align}
 (\Box_{\subE}-m^{2}-\xi\,R)\phi=0,
\end{align}
where $\Box_{\subE}$ is the d'Alembertian operator with respect to the Euclidean metric, $m$ is the scalar field mass and $\xi$ is the constant that couples the scalar to the gravitational field. The corresponding Euclidean Green function has the following mode-sum representation
\begin{align}
	G(x,x')=\frac{\kappa}{2\pi}\sum_{n=-\infty}^{\infty}e^{i n \kappa \Delta\tau}\sum_{l=0}^{\infty}\frac{(l+\mu)}{\mu\,\Omega_{d-2}}C_{l}^{\mu}(\cos\gamma)g_{nl}(r,r')
\end{align}
where $\mu=(d-3)/2$ and $\Omega_{d-2}=2\,\pi^{\mu+1}/\Gamma(\mu+1)$, $C_{l}^{\mu}(x)$ is the Gegenbauer polynomial and $\gamma$ is the geodesic distance on the $(d-2)$-sphere. The radial Green function satisfies
\begin{align}
	\label{eq:radialeqn}
	\Bigg[\frac{d}{dr}\Big(r^{d-2}f(r)\frac{d}{dr}\Big)-r^{d-2}\Big(\frac{n^{2}\kappa^{2}}{f(r)}+m^{2}+\xi\,R(r)\Big)\nonumber\\
	-r^{d-4}l(l+d-3)\Bigg]g_{nl}(r,r')=-\delta(r-r').
\end{align}
The solution can be expressed as a normalized product of homogeneous solutions
\begin{align}
	\label{eq:radialgreenfn}
	g_{nl}(r,r')=N_{nl}\,p_{nl}(r_{<})q_{nl}(r_{>}),
\end{align}
where $p_{nl}(r)$ and $q_{nl}(r)$ are homogeneous solutions which are regular on the horizon and the outer boundary (usually spatial infinity), respectively. We have adopted the notation $r_{<}\equiv \min\{r,r'\}$, $r_{>}\equiv \max\{r,r'\}$. The normalization constant is given by
\begin{align}
	\label{eq:norm}
	N_{nl}\,W\{p_{nl}(r),q_{nl}(r)\}=-\frac{1}{r^{d-2}f(r)},
\end{align}
where $W\{p,q\}$ denotes the Wronskian of the two solutions.

\section{The Singular Propagator}
The point-splitting approach to computing the vacuum polarization in the Hartle-Hawking state involves taking the coincidence limit of the regularized Euclidean Green function. The Green function is regularized by subtracting an appropriate parametrix. The prescription is known to be ambiguous \cite{WaldQFT}, and different singular parametrices will lead to different expressions for the vacuum polarization. However, provided the parametrix is symmetric and depends only on the local geometry, then the difference between two regularization prescriptions is a regular scalar that depends only on the metric and its derivatives. Moreover, this ambiguity is degenerate with ambiguities in the renormalizations of coefficients of higher curvature terms in the semi-classical Einstein equations. This guarantees that the semi-classical equations are invariant under the choice of regularization prescription.

Here, we adopt the Hadamard regularization prescription (see, e.g., \cite{DecaniniFolacciHadamardRen}), i.e., we define our singular propagator to be a Hadamard parametrix. In odd dimensions, we choose
\begin{align}
	G_{\subS}(x,x')=\frac{\Gamma(\frac{d}{2}-1)}{2 (2\pi)^{d/2}}\frac{U(x,x')}{\sigma(x,x')^{\frac{d}{2}-1}}.
\end{align}
The biscalar $\sigma(x,x')$ is the world function with respect to the Euclideanized metric. The biscalar $U(x,x')$ is smooth and symmetric in its arguments. For a scalar field, $U(x,x')$ satisfies the wave equation
\begin{align}
	\sigma(\Box-m^{2}-\xi\,R)U&=(d-2)\sigma^{a}\nabla_{a}U\nonumber\\
	&-(d-2)U\,\Delta^{-1/2}\sigma^{a}\nabla_{a}\Delta^{1/2},
\end{align}
where $\sigma^{a}\equiv \nabla^{a}\sigma$ and $\Delta(x,x')$ is the Van Vleck Morrette determinant. Assuming the Hadamard ansatz for a series solution
\begin{align}
	U(x,x')=\sum_{p=0}^{\infty}U_{p}(x,x')\,\sigma^{p},
\end{align}
it can be shown that each coefficient $U_{p}(x,x')$ satisfies
\begin{align}
	\label{eq:HadamardUp}
	&(p+1)(2p+4-d)U_{p+1}+(2p+4-d)\sigma^{a}\nabla_{a}U_{p+1}\nonumber\\
	&-(2p+4-d)U_{p}\Delta^{-1/2}\sigma^{a}\nabla_{a}\Delta^{1/2}\nonumber\\
	&+(\Box-m^{2}-\xi\,R)U_{p}=0,
\end{align}
with boundary condition $U_{0}=\Delta^{1/2}$.

The world function possesses a standard coordinate expansion which to lowest order is simply $\sigma=\tfrac{1}{2}g_{ab}\Delta x^{a}\Delta x^{b}+\textrm{O}(\Delta x^{3})$. In our first departure from the standard approach, we shall eschew the usual coordinate expansions and instead assume an expansion of the form
\begin{align}
	\label{eq:SigmaExp}
	\sigma=\sum_{ijk}\sigma_{ijk}(r)w^{i}\Delta r^{j}s^{k}
\end{align}
where 
\begin{align}
	\label{eq:ExpVar}
	w^{2}=\frac{2}{\kappa^{2}}(1-\cos \kappa\Delta\tau),\qquad s^{2}=f(r)\,w^{2}+2 r^{2}(1-\cos\gamma).
\end{align}
We will formally treat $w$ and $s$ as $\textrm{O}(\epsilon)\sim\textrm{O}(\Delta x)$ quantities. Substituting this into the defining equation for $\sigma$ and equating order by order uniquely determines the coefficients $\sigma_{ijk}(r)$. To leading order, we simply have $\sigma=\tfrac{1}{2}(s^{2}+\Delta r^{2}/f)+\textrm{O}(\epsilon^{3})$. An analogous expansion may be assumed for $U_{p}(x, x')$,
\begin{align}
	\label{eq:UExp}
	U_{p}(x,x')=\sum_{ijk}u^{(p)}_{ijk}(r)w^{i}\Delta r^{j}s^{k},
\end{align}
and substituting this into (\ref{eq:HadamardUp}) determines the coefficients $u^{(p)}_{ijk}(r)$.

Combining (\ref{eq:SigmaExp}) and (\ref{eq:UExp}) gives a series expansion for the Hadamard parametrix in terms of the expansion parameters $w$, $s$ and $\Delta r$. This type of computation is ideally suited to a symbolic computer package such as Mathematica. Since we are ultimately interested in the coincidence limit, let us simplify by taking the partial coincidence limit $\Delta r=0$, then it can be shown that $U/\sigma^{d/2-1}$ possesses an expansion of the form
\begin{align}
	\frac{U}{\sigma^{\frac{d}{2}-1}}=\sum_{i=0}^{ \mu+m}\sum_{j=-i}^{ i}\mathcal{D}_{ij}(r)\epsilon^{2i-2\mu-1}\frac{w^{2i+2j}}{s^{2\mu+2j+1}}+\textrm{O}(\epsilon^{2m+1}).
\end{align}
The coefficients $\mathcal{D}_{ij}(r)$ for the $d=5$ Schwarzschild-Tangherlini spacetime is given in Table \ref{tab:Dcoeff}. For higher odd dimensions, the expressions are too large to be useful in print-form, however, a Mathematica Notebook containing the expressions is available online \cite{MyWebpage}. We could truncate this sum at $i=\mu$ since higher-order terms vanish in the coincidence limit. However, it will be useful later to keep terms at least up to $\textrm{O}(\epsilon)$, the higher-order terms will speed the convergence of the mode-sum expression for the regularized Green function. Let us take $m=1$ and separate out the negative $j$ terms to get
\begin{align}
	\label{eq:Hadamardclosed}
	\frac{U}{\sigma^{\frac{d}{2}-1}}&=\sum_{i=0}^{ \mu+1}\sum_{j=0}^{ i}\mathcal{D}_{ij}(r)\,\epsilon^{2i-2\mu-1}\frac{w^{2i+2j}}{s^{2\mu+2j+1}}\nonumber\\
	&+\sum_{i=1}^{ \mu+1}\sum_{j=1}^{ i}\mathcal{D}_{i,-j}(r)\,\epsilon^{2i-2\mu-1}\frac{w^{2i-2j}}{s^{2\mu-2j+1}}\nonumber\\
	&+\textrm{O}(\epsilon^{3}).
\end{align}
This is our second major departure from the usual treatment; we have point-split in multiple directions. It seems natural to avail of our freedom to point-split in any direction to choose a splitting only in one direction, and choosing that direction to be along a Killing vector seems to greatly simplify the expressions for the parametrix. However, employing this freedom too early is actually a hindrance since what is actually needed is a mode-sum expression for the parametrix, not a closed-form expression. A mode-sum expression is most naturally obtained by a simultaneous Fourier and multipole decomposition of the parametrix, which requires splitting in both the temporal and angular directions. The mode-sum decomposition is explicitly derived in the next section.

\section{Mode-Sum Representation of the Hadamard Parametrix}
We wish to decompose the terms of the form $w^{2i\pm 2j}/s^{2\mu\pm 2j+1}$ in terms of Fourier frequency modes and multipole moments. If this can be achieved then a mode-by-mode subtraction for the regularized Green function is feasible. Start by writing
\begin{align}
	\frac{w^{2i\pm 2j}}{s^{2\mu\pm 2j+1}}=\sum_{n=-\infty}^{\infty}e^{in\kappa\Delta\tau}\sum_{l=0}^{\infty}(2 l+2\mu)C_{l}^{\mu}(\cos\gamma)\nonumber\\
	\times\,\,\stackrel{[d]}{\Psi}\!\!{}_{nl}(i,\pm j|r).
\end{align}
The task is to determine the regularization parameters $\stackrel{[d]}{\Psi}\!\!{}_{nl}(i,\pm j|r)$. With $x=\cos\gamma$, multiplying both sides by $e^{-i n'\Delta \tau}(1-x^{2})^{\mu-\frac{1}{2}}C_{l'}^{\mu}(x)$ and integrating gives
\begin{align}
	\stackrel{[d]}{\Psi}\!\!{}_{nl}(i,\pm j|r)=\frac{\kappa}{(2\pi)^{2}}\frac{2^{2\mu-1}\Gamma(\mu)^{2}l!}{\Gamma(l+2\mu)}\int_{0}^{2\pi/\kappa}\int_{-1}^{1}\frac{w^{2i\pm 2j}}{s^{2\mu\pm 2j+1}}\nonumber\\
	\times\,\,e^{-i n \kappa\Delta\tau}(1-x^{2})^{\mu-\frac{1}{2}}C_{l}^{\mu}(x)\,dx\,d\Delta\tau,
\end{align}
where we have used the completeness relations
\begin{align}
	&\int_{0}^{2\pi/\kappa}e^{-i (n-n')\Delta\tau}d\Delta\tau=\frac{2\pi}{\kappa}\delta_{n n'},\nonumber\\
	&\int_{-1}^{1}(1-x^{2})^{\mu-\frac{1}{2}}C_{l}^{\mu}(x)C_{l'}^{\mu}(x)\,dx=\frac{2^{1-2\mu}\pi\,\Gamma(n+2\mu)}{(l+\mu)\,l!\,\Gamma(\mu)^{2}}\delta_{l l'}.
\end{align}
We perform the $x$ integration above by employing the identity \cite{CohlGegenInt}
\begin{align}
&\int_{-1}^{1}\frac{(1-x^{2})^{\mu-1/2}C_{l}^{\mu}(x)}{(z-x)^{\mu\pm j+1/2}}dx\nonumber\\
&=\frac{(-1)^{j}\sqrt{\pi}\Gamma(l+2\mu)(z^{2}-1)^{\mp j/2}}{2^{\mu-3/2}l!\Gamma(\mu)\Gamma(\mu\pm j+1/2)}Q^{\pm j}_{l+\mu-1/2}(z),	
\end{align}
to obtain
\begin{align}
	\label{eq:RegParamInt}
	&\stackrel{[d]}{\Psi}\!\!{}_{nl}(i, \pm j|r)=\frac{\kappa}{(2\pi)^{2}}\frac{2^{i}\sqrt{\pi}(-1)^{j}\Gamma(\mu)}{\kappa^{2i \pm 2j}r^{2\mu\pm 2j+1}\Gamma(\mu+\frac{1}{2}\pm j)}\nonumber\\
	&\times\,\int_{0}^{2\pi/\kappa}(1-\cos\kappa t)^{i\pm j}e^{-i n \kappa t}(z^{2}-1)^{\mp j/2}Q^{\pm  j}_{l+\mu-\frac{1}{2}}(z) dt,
\end{align}
with
\begin{align}
	z=1+\frac{f^{2}}{\kappa^{2}r^{2}}(1-\cos\kappa t).
\end{align}
We note that in odd $d>3$, the parameter $\mu=(d-3)/2$ is always a positive integer. In particular, we note that since $l+\mu\pm j-1/2$ is not a negative integer, the associated Legendre function of the second kind appearing in the integral representation of the regularization parameters above is always well-defined.

We will compute the $\stackrel{[d]}{\Psi}\!\!{}_{nl}(i, j|r)$ terms first. Using the fact that
\begin{align}
	\label{eq:LegDeriv}
	(z^{2}-1)^{-j/2}Q_{\nu}^{j}(z)=\frac{(-1)^{j}}{2^{j}(1-\cos\kappa t)^{j}}\Big(\frac{1}{\eta}\frac{\partial}{\partial\eta}\Big)^{j}Q_{\nu}(z),
\end{align}
where
\begin{align}
	\eta\equiv \sqrt{1+\frac{f(r)}{\kappa^{2}r^{2}}},
\end{align}
we arrive at
\begin{align}
	&\stackrel{[d]}{\Psi}\!\!{}_{nl}(i,j|r)=\frac{\kappa}{(2\pi)^{2}}\frac{2^{i-j}\sqrt{\pi}\Gamma(\mu)}{\kappa^{2i \pm 2j}r^{2\mu+ 2j+1}\Gamma(\mu+\frac{1}{2}+j)}\nonumber\\
	&\times\,\Bigg(\frac{1}{\eta}\frac{\partial}{\partial\eta}\Bigg)^{j}\int_{0}^{2\pi/\kappa}(1-\cos\kappa t)^{i}e^{-i n \kappa t}Q_{l+\mu-\frac{1}{2}}(z) dt.
\end{align}
In order to perform the integral we must factor out the time dependence from the Legendre function, which may be achieved by employing the addition theorem \cite{GradRiz},
\begin{align}
	\label{eq:AddThm}
	Q_{\nu}(z)=P_{\nu}(\eta)Q_{\nu}(\eta)+2\sum_{p=1}^{\infty}(-1)^{p}P_{\nu}^{-p}(\eta)Q_{\nu}^{p}(\eta)\cos p\,\kappa t,
\end{align}
whence the time integral reduces to
\begin{align}
	\label{eq:timeint}
	&\int_{0}^{2\pi/\kappa}(1-\cos\kappa t)^{i}e^{-i n\kappa t}\cos p \kappa t\,dt=\nonumber\\
	&\frac{\sqrt{\pi }}{\kappa}\Bigg[ \frac{2^i i! \Gamma \left(i+\frac{1}{2}\right) (-1)^{n-p}}{(i+p-n)! (i-p+n)!}+\frac{2^i
   i! \Gamma \left(i+\frac{1}{2}\right) (-1)^{p+n}}{(i-p-n)! (i+p+n)!}\Bigg].
\end{align}
The factorials in the denominator imply that that there is a finite number of integer $p$ for which the integral is nonzero. In particular, the first term on the right-hand side of (\ref{eq:timeint}) is nonzero only for $|p-n|\le i$ while the second term is nonzero for $|p+n|\le i$. The range is further restricted in our case since $p\ge 1$ and hence the sets of integers $p$ for which the first and second terms are nonzero are $p\in \{\max(1,n-i),n+i\}$ and $p\in\{\max(1, -n-i),i-n\}$, respectively. An equivalent expression for (\ref{eq:timeint}) in terms of a sum of Kronecker deltas is easily derived. Putting these together, we obtain
\begin{widetext}
\begin{align}
	\label{eq:RegParamP}
\stackrel{[d]}{\Psi}\!\!{}_{nl}(i,j|r)&=\frac{2^{2i-j-1}(-1)^{n}i!\,\Gamma(i+\frac{1}{2})\Gamma(\mu)}{\pi\kappa^{2i + 2j}r^{2\mu+ 2j+1}\Gamma(j+\mu+\tfrac{1}{2})}\left(\frac{1}{\eta}\frac{d}{d \eta}\right)^{j}\Bigg\{\frac{P_{l+\mu-\frac{1}{2}}(\eta)Q_{l+\mu-\frac{1}{2}}(\eta)}{(i-n)!(i+n)!}\nonumber\\
&+\sum_{p=\max\{1,n-i\}}^{i+n}\frac{P^{-p}_{l+\mu-\frac{1}{2}}(\eta)Q^{p}_{l+\mu-\frac{1}{2}}(\eta)}{(i+p-n)!(i-p+n)!}+\sum_{p=\max\{1,-n-i\}}^{i-n}\frac{P^{-p}_{l+\mu-\frac{1}{2}}(\eta)Q^{p}_{l+\mu-\frac{1}{2}}(\eta)}{(i+p+n)!(i-p-n)!}\Bigg\}.
\end{align}

To derive the $\stackrel{[d]}{\Psi}\!\!{}_{nl}(i, -j|r)$ terms, we make use of the following result (this result may be new, we did not find it in any of the standard references on Legendre functions; it is straightforward to prove by induction)
\begin{align}
(z^{2}-1)^{j/2}Q^{-j}_{\nu}(z)=\sum_{k=0}^{j}\frac{(-1)^{k}}{2^{j+1}}\binom{j}{k}\frac{(2\nu+2 j-4k+1)}{\prod_{q=0}^{j}(\nu-k+\tfrac{1}{2}+q)}Q_{\nu+j-2k}(z).
\end{align}
When $\nu-k+\tfrac{1}{2}> 0$, we can simplify by replacing the product with its Pocchammer representation  $\prod_{q=0}^{j}(\nu-k+\tfrac{1}{2}+q)=(\nu-k+\frac{1}{2})_{j+1}$. Employing this identity in (\ref{eq:RegParamInt}) gives
\begin{align}
	\stackrel{[d]}{\Psi}\!\!{}_{nl}(i, - j|r)=\frac{\kappa}{(2\pi)^{2}}\frac{2^{i-j}\sqrt{\pi}(-1)^{j}\Gamma(\mu)}{\kappa^{2i - 2j}r^{2\mu- 2j+1}\Gamma(\mu+\frac{1}{2}- j)}\sum_{k=0}^{j}(-1)^{k}\binom{j}{k}\frac{(l+\mu+j-2k)}{\prod_{p=0}^{j}(l+\mu+p-k)}\nonumber\\
	\times\,\int_{0}^{2\pi/\kappa}(1-\cos\kappa t)^{i- j}e^{-i n \kappa t}Q_{l+\mu-\frac{1}{2}+j-2k}(z) dt.
\end{align}
We now proceed as above: we apply the addition theorem (\ref{eq:AddThm}) to isolate the time-dependence, and integrate using (\ref{eq:timeint}). The result is
\begin{align}
	\label{eq:RegParamN}
\stackrel{[d]}{\Psi}\!\!{}_{nl}(i,-j|r)&=\frac{2^{2i-2j-1}(-1)^{n+j}(i-j)!\Gamma(i-j+\tfrac{1}{2})\Gamma(\mu)}{\pi\kappa^{2i - 2j}r^{2\mu- 2j+1}\Gamma(\mu+\tfrac{1}{2}-j)}\sum_{k=0}^{j}(-1)^{k}\binom{j}{k}\frac{(l+\mu+j-2k)}{\prod_{q=0}^{j}(l+\mu+q-k)}\nonumber\\
&\times\,\,\Bigg\{\frac{P_{l+\mu-\frac{1}{2}+j-2k}(\eta)Q_{l+\mu-\frac{1}{2}+j-2k}(\eta)}{(i-j-n)!(i-j+n)!}+\sum_{p=\max\{1,n-i+j\}}^{i-j+n}\frac{P^{-p}_{l+\mu-\frac{1}{2}+j-2k}(\eta)Q^{p}_{l+\mu-\frac{1}{2}+j-2k}(\eta)}{(i-j+p-n)!(i-j-p+n)!}\nonumber\\
&+\sum_{p=\max\{1,-n-i+j\}}^{i-j-n}\frac{P^{-p}_{l+\mu-\frac{1}{2}+j-2k}(\eta)Q^{p}_{l+\mu-\frac{1}{2}+j-2k}(\eta)}{(i-j+p+n)!(i-j-p-n)!}\Bigg\}.
\end{align}
\end{widetext}
Eqs. (\ref{eq:RegParamP}) and (\ref{eq:RegParamN}) are the regularization parameters for a scalar field in a static spherically symmetric spacetime in arbitrary dimensions. On the one hand, these expressions look complicated, however, it is remarkably elegant that all the regularization parameters in a static, spherically symmetric spacetime in arbitrary odd dimensions can be written as a finite sum of products of associated Legendre functions. In terms of these regularization parameters, the mode-sum representation of the singular field is
\begin{align}
	G_{\subS}(x,x')=&\frac{\Gamma(\frac{d}{2}-1)}{2(2\pi)^{d/2}}\sum_{l=0}^{\infty}(2l+2\mu)C_{l}^{\mu}(\cos\gamma)\sum_{n=-\infty}^{\infty}e^{in\kappa\Delta\tau}\nonumber\\
	&\times\,\,\Big\{\sum_{i=0}^{\mu+1}\sum_{j=0}^{i}\mathcal{D}_{ij}(r)\stackrel{[d]}{\Psi}\!\!{}_{nl}(i,j|r)\nonumber\\
	&+\sum_{i=1}^{\mu+1}\sum_{j=1}^{i}\mathcal{D}_{i,-j}(r)\stackrel{[d]}{\Psi}\!\!{}_{nl}(i,-j|r)\Big\}+\textrm{O}(\epsilon^{3}).
\end{align}
This is the main result. It allows one to numerically compute the regularized vacuum polarization in arbitrary odd dimensions in an extremely efficient way. We describe this calculation for a massless scalar field in the Schwarzschild-Tangherlini spacetimes in the following section.

\section{Vacuum Polarization in Schwarzschild-Tangherlini Spacetime}
In this section we outline the numerical implementation of the reguarization scheme described above, applied to a massless scalar field in the higher-dimensional generalizations of the Schwarzschild black hole: the Schwarzschild-Tangherlini spacetimes. We note that the restriction to massless fields is a minor convenience, it is completely straight-forward to generalize this computation to massive fields. In particular, the regularization parameters $\stackrel{[d]}{\Psi}\!\!{}_{nl}(i,j|r)$ derived above do not depend on the mass, but only on the local geometry of the spacetime. The mass eneters into the calculation through the coefficients $\mathcal{D}_{ij}(r)$.

Now in the usual Schwarzschild coordinates, the Schwarzschild-Tangherlini metric takes the form (\ref{eq:metric}) with
\begin{align}
	\label{eq:fST}
	f(r)=1-\Big(\frac{r_{\textrm{h}}}{r}\Big)^{d-3}.
\end{align}
These coordinates are singular at $r=r_{\textrm{h}}$ which corresponds to the black hole horizon. For simplicity, throughout the remainder of this section, we work in units where $r_{\textrm{h}}=1$. That implies that the surface gravity $\kappa=\tfrac{1}{2}f'(r_{\textrm{h}})=\tfrac{1}{2}(d-3)$.

\subsection{Calculation of radial modes}
We briefly describe our numerical computation of the radial modes $p_{nl}(r)$ and $q_{nl}(r)$, the homogeneous solutions to (\ref{eq:radialeqn}) which are regular on the horizon and at $\infty$ respectively. For $f(r)$ given by (\ref{eq:fST}), solutions cannot in general be given in terms of known functions and must be solved numerically. However, for $n=0$ this equation reduces to
\begin{align}
\label{eqn:hom}
&\left[\frac{d}{dr} (r^{d-2}-r)\frac{d }{dr} -r^{d-4}l(l+d-3)\bigg)\right] S=0
\end{align}
which possesses solutions in terms of Legendre functions. To see this, we introduce a new independent variable $x=2r^{d-3}-1$. Then Eqn. (\ref{eqn:hom}) takes the form
\begin{align*}
&\left[\frac{d}{d x} (x^2-1)\frac{d }{d x} -L(L+1)\bigg)\right] R(x)=0
\end{align*}
where $L=l/(d-3)$. This is now in the form of Legendre's differential equation which possesses the following pair of independent solutions:
\bea
\label{eqn:n0}
p_{0l}(r)&=P_{l/(d-3)} (2r^{d-3}-1)\\
q_{0l}(r)&=Q_{l/(d-3)} (2r^{d-3}-1)
\eea
where $P_{\nu}(z)$ and $Q_{\nu}(z)$ are Legendre functions of the first and second kind, respectively. The Wronskian of the $n=0$ modes is
\bes
W\{p_{0l}(r),q_{0l}(r)\}= -\frac{d-3}{2(r^{d-2}-r)}.
\ees
Comparing this with (\ref{eq:norm}) we see that
\bes
N_{0l}=\frac{2}{d-3}=\frac{1}{\kappa}.
\ees

To calculate $p_{nl}(r)$ for $n\ne 0$, we first note that the solutions are invariant under $n\to-n$. Hence, we need only consider positive frequency modes. To compute $p_{nl}(r)$ for $n>1$, we integrate the homogeneous version of (\ref{eq:radialeqn}) from an initial point near the horizon outwards. Employing a standard Frobenius analysis, a series solution about the regular singular point at $r=r_{\textrm{h}}=1$ is obtained and used as a starting value for the integration of Eq.~(\ref{eq:radialeqn}). We begin our numerical integration at a distance of $1/1000$ from the horizon and the upper limit of integration is set at a distance of $r=21$. Given that we expect the majority of the interesting features in our results to be in the vicinity of the black hole horizon, we choose to calculate on a ``tortoise coordinate''-like grid, where the sampling points are more dense in the near-horizon region. We chose to perform the numerical integration using the NDSolve package in Mathematica, which calculates $p_{nl}(r)$ on a grid and interpolates between each grid point. 

\begin{figure}
\includegraphics[width=0.95\columnwidth]{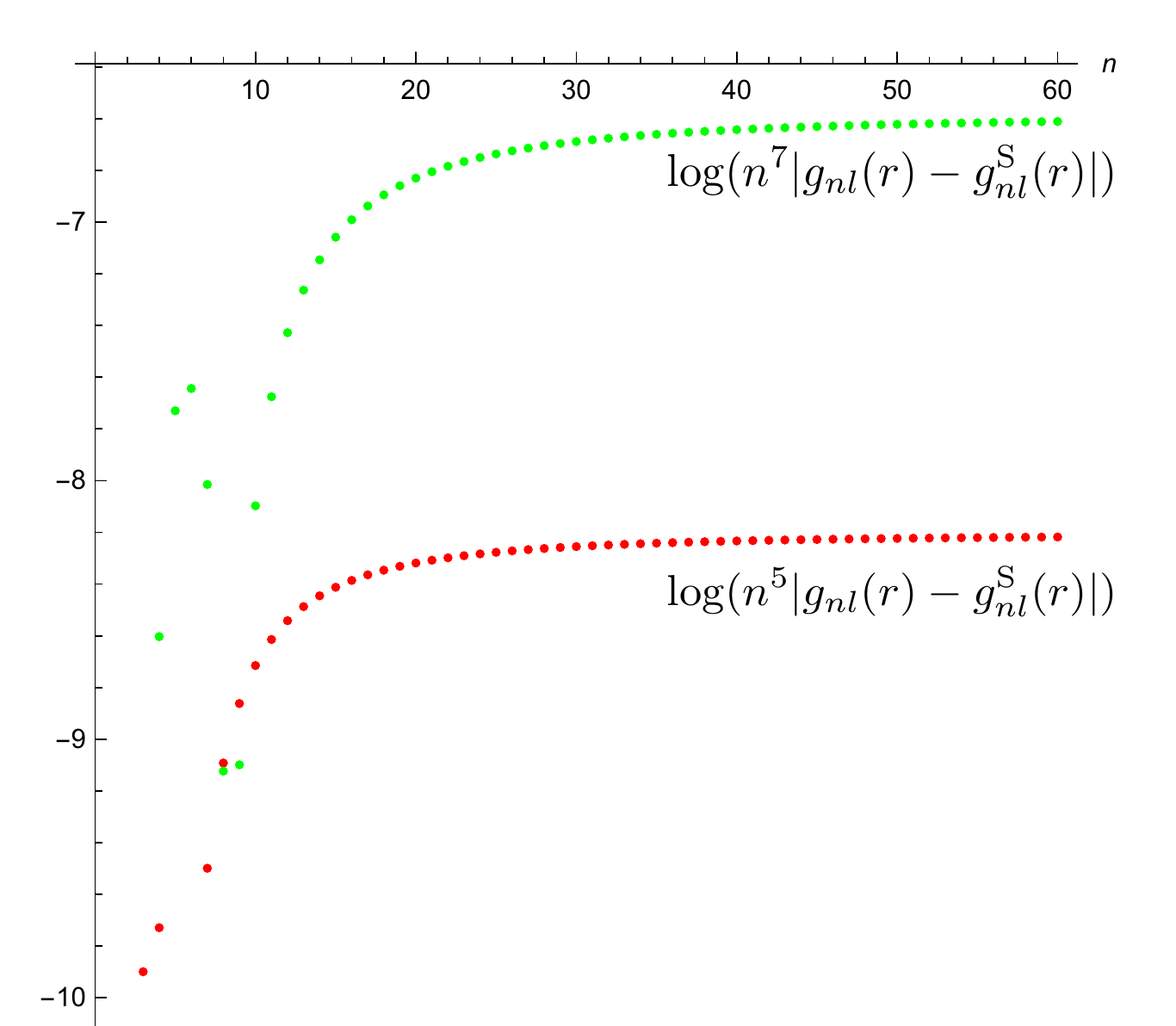}
\caption{Log plots showing convergence over $n$ in the mode sums expression. The red line represents $\log(n^5 |g_{nl}(r)-g_{nl}^{\subS}(r)|)$ where we do not include the $\textrm{O}(\epsilon)$ terms in $g_{nl}^{\subS}(r)$. The plot shows that the difference $g_{nl}(r)-g_{nl}^{\subS}(r)$ scales like $n^{-5}$ for large $n$. The green line represents $\log(n^7|g_{nl}(r)-g_{nl}^{\subS}(r)|)$ where we have included $\textrm{O}(\epsilon)$ terms in the singular summand. The plot shows that the difference scales like $n^{-7}$ for large $n$. }
\label{fig:d5nconvergence}
\end{figure}

Once we have calculated $p_{nl}(r)$, we may then obtain the other solution $q_{nl}(r)$ by integrating the Wronskian relation (\ref{eq:norm}) between $r$ and $\infty$. This leads to the following expression for $q_{nl}(r)$
\be
q_{nl}(r)=\frac{p_{nl}(r)}{N_{nl}} \int_{r}^{\infty}\frac{dr'}{r'^{d-2}f(r')(p_{nl}(r'))^2}
\ee
In practice we must set the upper limit of integration to be at a finite large value, we set this to be the end point of our integration for $p_{nl}(r)$. Since $p_{nl}(r)$ is growing exponentially with $r$, the errors incurred by truncating the integral at this point will be negligible. It is worth noting here that we could also have obtained $q_{nl}(r)$ by integrating the homogeneous version of (\ref{eq:radialeqn}) inwards from some large $r$-value. In this case an initial value would be calculated by constructing an asymptotic series about $r=\infty$, which is an irregular singular point of the equation. However integrating the Wronskian condition appears to lead to more accurate results than integrating the differential equation. Finally in the numerical calculation of both $p_{nl}(r)$ and $q_{nl}(r)$ the internal working precision of each calculation is set to 50 digits while the accuracy and precision goals (i.e. the effective number of digits of precision and accuracy sought in the final result) were both set to 35 digits.

\subsection{Mode-Sum Calculation}
Armed with an accurate numerical evaluation of the radial Green function and explicit closed-form expressions for the regularization parameters, we are now in a position to calculate the vacuum polarization $\langle \phi^2 \rangle_{\textrm{ren}}$ for $d=5,7, 9$ and $11$. Let us simplify the notation by writing
\begin{align}
	G_{\subS}(x,x')=\frac{\kappa}{2\pi}\sum_{l=0}^{\infty}\frac{(l+\mu)}{\mu\,\Omega_{d-2}}C_{l}^{\mu}(\cos\gamma)\Big\{g_{0l}^{\subS}(r)\nonumber\\
	+2\sum_{n=1}^{\infty}\cos\kappa\Delta\tau\,g_{nl}^{\subS}(r)\Big\},
\end{align}
where
\begin{align}
	g_{nl}^{\subS}(r)=\frac{\mu\,\Omega_{d-2}}{\kappa}\frac{\Gamma(\frac{d}{2}-1)}{(2\pi)^{\frac{d}{2}-1}}\Big\{\sum_{i=0}^{\mu+1}\sum_{j=0}^{i}\mathcal{D}_{ij}(r)\stackrel{[d]}{\Psi}\!\!{}_{nl}(i,j|r)\nonumber\\
	+\sum_{i=1}^{\mu+1}\sum_{j=1}^{i}\mathcal{D}_{i,-j}(r)\stackrel{[d]}{\Psi}\!\!{}_{nl}(i,-j|r)\Big\}.
\end{align}
The Gegenbauer polynomial evaluated at coincidence are
\begin{align}
	C_{l}^{\mu}(1)=\binom{2\mu+l-1}{l},
\end{align}
and hence the vacuum polarization is given by
\begin{align}
\langle \phi^{2} \rangle_{\textrm{ren}}&=\lim_{x'\to x}\Big\{G(x,x')-G_{\subS}(x,x')\Big\}\nonumber\\
&=\frac{\kappa}{2\pi}\sum_{l=0}^{\infty}\frac{(l+\mu)}{\mu\,\Omega_{d-2}}\binom{2\mu+l-1}{l}\Big\{g_{0l}(r)-g_{0l}^{\subS}(r)\nonumber\\
&	+2\sum_{n=1}^{\infty}(g_{nl}(r)-g_{nl}^{\subS}(r))\Big\}.
\end{align}

The order in which the sums are performed here is dictated by the order in which the limits were taken (see  \cite{OttewillTaylorCS} for a discussion of this point). In our case, the temporal points were necessarily taken to coincidence before the angular points--this is simply a consequence of our definitions for expansion variables $s$ and $w$--and this implies that the $n$-sum must be performed first. The convergence of the inner sum over $n$ can be shown numerically to be $\textrm{O}(n^{-d-2})$ for each value of $d$ under consideration in this paper (see Fig \ref{fig:d5nconvergence} for plots of convergence for $d=5$).

\begin{figure}
	\includegraphics[width=9cm]{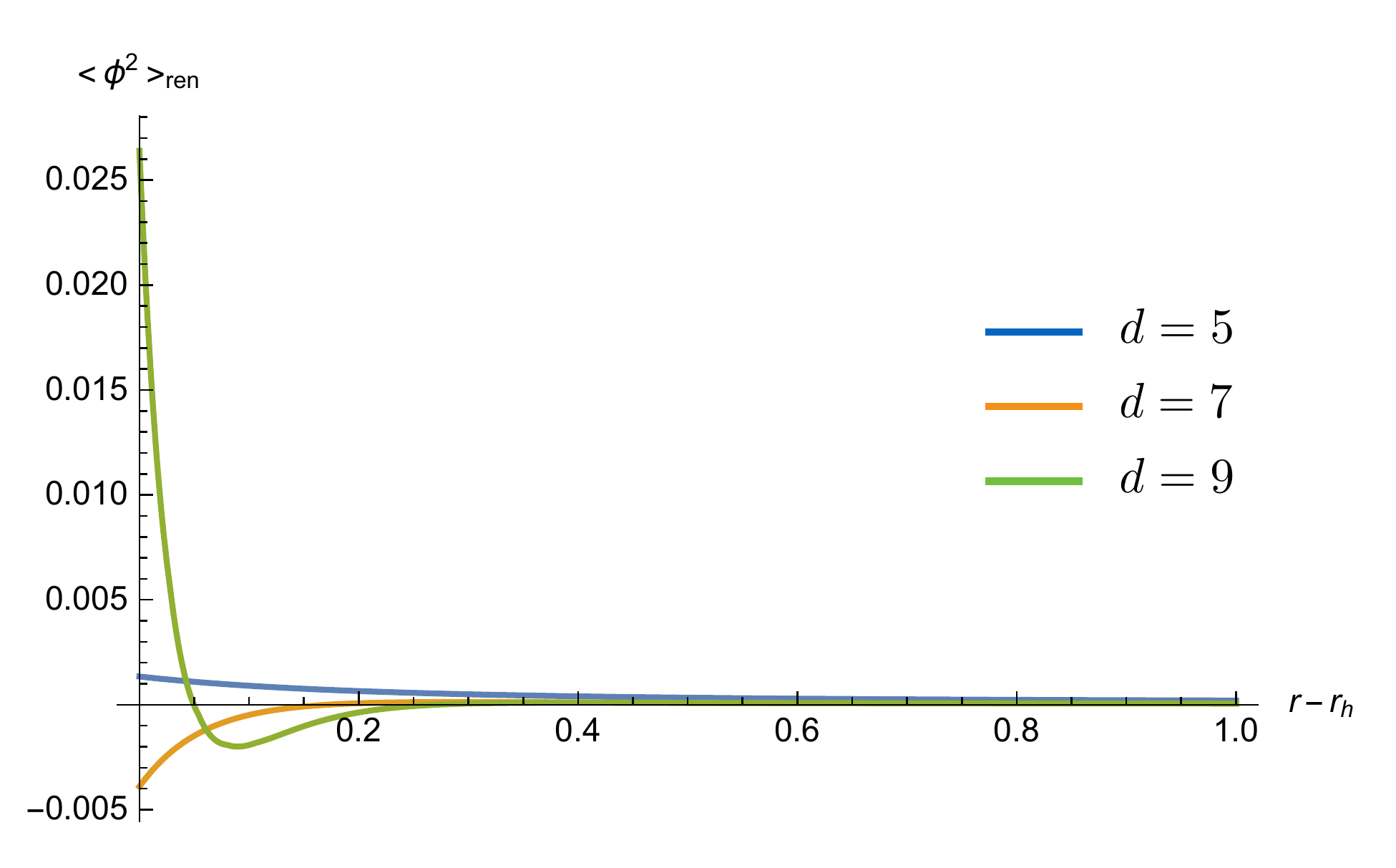}
	\caption{Plot of the renormalized vacuum polarization in the exterior region of a Schwarzschild-Tangherlini black hole as a function of the radial co-ordinate $r$ for spacetime dimensions $d=5,7$ and $d=9$. The event horizon is located at $r=r_{\textrm{h}}=1$.}
	\label{fig:phid579}
\end{figure}

\begin{figure*}
	\includegraphics[width=16cm]{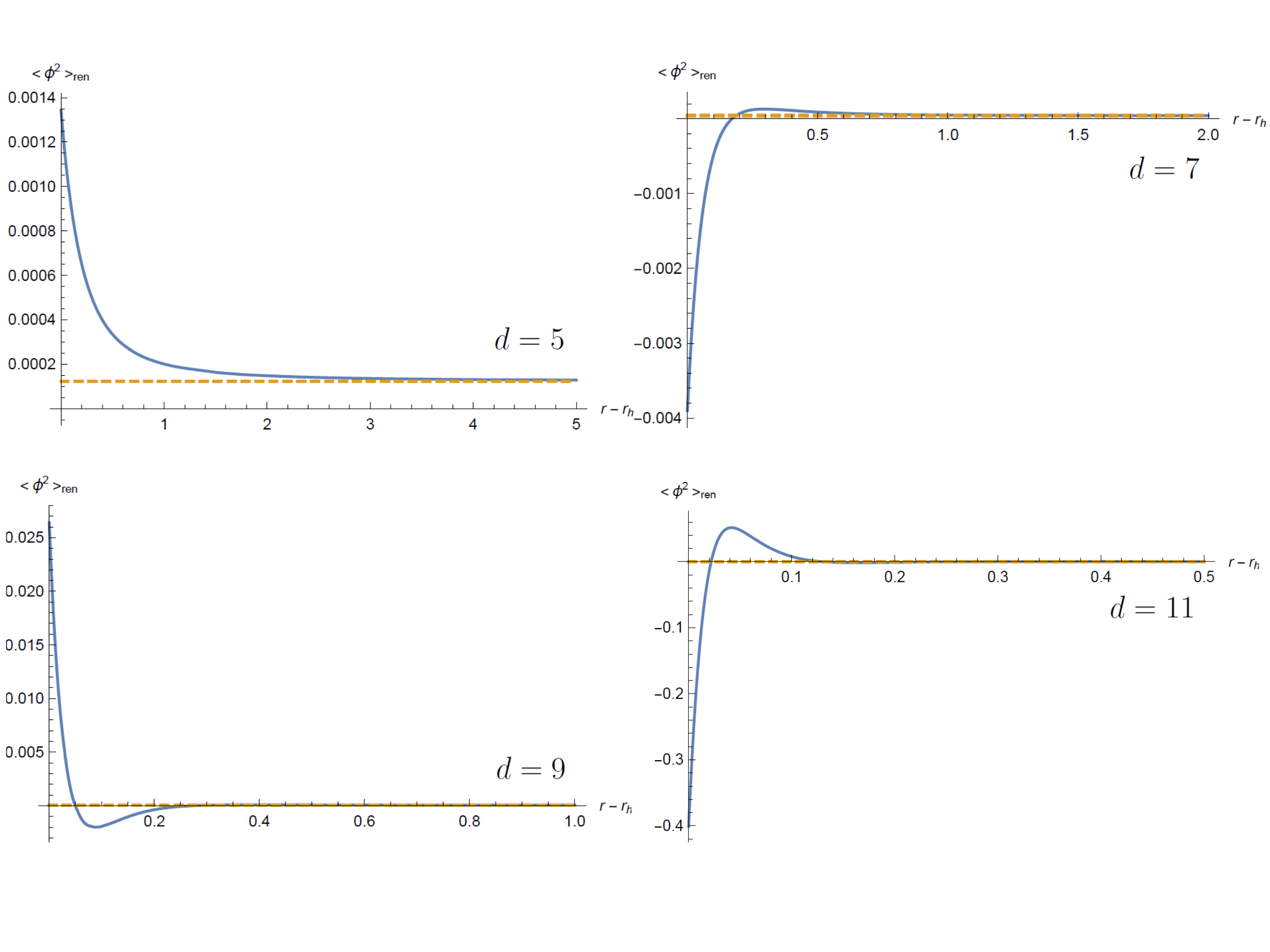}
	\caption{Plot of the renormalized vacuum polarization in the exterior region of a Schwarzschild-Tangherlini black hole as a function of the radial coordinate $r$ in various odd dimensions, from $d=5,..,11$. The event horizon is located at $r=r_{\textrm{h}}=1$. The dashed line is the asymptotic value given by Eq. (\ref{eq:flatphisq}).}
	\label{fig:phid11}
\end{figure*}

We present plots of $\langle \phi^2 \rangle_{\textrm{ren}}$ in the exterior region of a Schwarzschild-Tangherlini black hole space-time for $d=5,7,9,11$. In units where the black hole event horizon has been set to unity, the near-horizon vacuum polarization increases rapidly with number of dimensions. Hence, in Fig. \ref{fig:phid579} we present on the same graph the results for $d=5,7,9$; we exclude $d=11$ as its features dominate over the results from the other dimensions. This is followed by a series of individual plots for each dimension, in Fig.~\ref{fig:phid11}. From the plots, we might conjecture  that for $d=7, 11,...$, the vacuum polarization is rapidly increasing from the horizon out to some turning point, before decreasing and eventually approaching its value at infinity. For the alternate odd dimensions $9,13,...$, the vacuum polarization decreases rapidly from the horizon to some turning point, before slowly increasing and eventually asymptoting to its value at infinity. Moreover, the rate of change seems to be greater  and the turning point closer to the horizon as the number of dimensions is increased, though these may be artefacts of the units in which we are working.

As the Schwarzchild-Tangherlini spacetime is asymptotically flat we would expect that, as $r \to \infty$,  $\langle \phi^2 \rangle_{\textrm{ren}}$ would approach the value of the renormalised vacuum polarization for a scalar field at the Hawking temperature in flat spacetime. In addition, given the form of the metric function Eq. (\ref{eq:fST}) we would also expect the rate at which this occurs to increase with the number of spacetime dimensions.

The vacuum polarization for a massless thermal field at temperature $T$ propagating in a $d$-dimensional Minkowski spacetime can be computed in closed form (see, e.g., \cite{Hewitt:15})
\begin{equation}
	\label{eq:flatphisq}
\langle \phi^2 \rangle^{\mathcal{M}}_{\textrm{ren}} = \frac{\Gamma\left(\frac{d}{2}-1\right)T^{d-2}}{2\pi^{d/2}} \zeta(d-2)
\end{equation}
where $\zeta(x)$ is the Riemann zeta function. By setting $T$ equal to the Hawking temperature $T=\kappa/2\pi$ we can explicitly show in Fig. \ref{fig:phid11} that the renormalised vacuum polarization for a massless field in the Schwarzchild-Tangherlini spacetime with odd $d=5,...,11$ does indeed approach the flat spacetime value given by (\ref{eq:flatphisq}) and moreover we see that the rate at which this occurs increases with $d$.

It should be noted here that for the calculation of  $\langle \phi^2 \rangle_{\textrm{ren}}$, the first grid point is taken to be the value of $\langle \phi^2 \rangle_{\textrm{ren}}$ on the black hole horizon, calculated by extending the work of \cite{Fro:89}. The relevant horizon values for this paper are given in the table below.
{\renewcommand{\arraystretch}{1.2}
\begin{table}[!ht]
	\begin{tabular}{|c|c|}
		\hline
		$d$&$\langle \phi^2\rangle_{\textrm{ren}}$ at $r=r_{\textrm{h}}=1$\\ 
		\hline
		\rule{0pt}{4ex} 5&$\displaystyle{\frac{1}{24 \pi ^3}}$\\ 
		\rule{0pt}{4ex} 7&-$\displaystyle{\frac{11}{60 \pi^4}-\frac{1}{16 \pi^3}}$\\ 
		\rule{0pt}{4ex} 9&  0.02639370185\\
		\rule{0pt}{4ex} 11&-0.40082310320\\
		\hline
	\end{tabular}
	\label{tab:hor}
\end{table}}
While the result for $d=5$ was derived in \cite{Fro:89}, to the best of the authors' knowledge this is the first instance where analytical results for $d=7$ are given. The results for $d=9$ and $d=11$ were calculated numerically. It is worth noting here that in each of the below plots the value of $\langle \phi^2 \rangle_{\textrm{ren}}$ at the last numerically calculated grid point matches up smoothly with the horizon value. This demonstrates that the method developed in this paper is uniform across the entire exterior region. While previous uniform methods of calculating $\langle \phi^2 \rangle_{\textrm{ren}}$ using extended Green-Liouville asymptotics \cite{breen:10} have been developed for $d=4$, the majority of previous calculations have relied on the WKB approximation, which breaks down near the horizon. We feel that this uniformity is a major advantage of the method presented in this paper.

\section{Conclusions}
We have presented a new and systematic method for computing vacuum polarization in odd dimensions in static, spherically-symmetric spacetimes. The method departs from the usual approach in two significant ways: First, we expand the Hadamard parametrix in a judicious choice of variables, not in the usual coordinate separations. Second, we point-split in multiple directions. These two combined allow us to do a simultaneous decomposition of the Hadamard parametrix in Fourier frequency modes and multipole moments. In fact, the coefficients of this decomposition--which we call the regularization parameters--can be determined in closed-form in arbitrary dimensions. Our approach results in a mode-by-mode subtraction for the vacuum polarization that is rapidly converging--because higher-order terms in the singular parametrix are easy to include within this prescription--and hence straight-forward to numerically evaluate to high accuracy. Moreover, the resultant mode-sum enjoys a convergence that is uniform in the distance from the horizon, a property that is not shared by methods based on the WKB approximation.

\begin{table*}
	\begin{center}
	\begin{tabular}{|c|ccccc|} \hline
		& \multicolumn{5}{c|}{$\mathcal{D}_{ij}(r)$ coefficients for 5D Schwarzschild-Tangherlini}\\ \hline \hline
		& & & & & \\ 
$\mathcal{D}_{00}$& $2 \sqrt{2}$ & \text{} & \text{} & \text{} & \text{} \\
$\mathcal{D}_{1j}$ for $j=-1,...,1$ & $-\frac{1}{2 \sqrt{2} r^4}$ & $\frac{\sqrt{2} \left(r^2-1\right)}{r^6}$ &
   $-\frac{\left(r^2-1\right)^2 \left(r^4+r^2+4\right)}{2 \sqrt{2} r^8}$ & \text{} &
   \text{} \\
$\mathcal{D}_{2j}$ for $j=-2,...,2$& $-\frac{16 r^2+29}{160 \sqrt{2} r^8}$ & $\frac{2 r^4-5 r^2+3}{4 \sqrt{2} r^{10}}$ &
   $\frac{\left(r^2-1\right)^2 \left(15 r^4-41 r^2+196\right)}{80 \sqrt{2} r^{12}}$ &
   $-\frac{\left(r^2-1\right)^3 \left(4 r^8+8 r^6+42 r^4+23 r^2+148\right)}{60 \sqrt{2}
   r^{14}}$ & $\frac{5 \left(r^2-1\right)^4 \left(r^4+r^2+4\right)^2}{96 \sqrt{2} r^{16}}$
   \\ & & & & &\\ \hline
   \end{tabular}
   \caption{We list the coefficients $\mathcal{D}_{ij}(r)$ for the $d=5$ Schwarzschild-Tangherlini spacetime. The horizon radius has been set to unity. These coefficients arise in the decomposition of the Hadamard parametrix in our variables $s$ and $w$ defined in Eq. \eqref{eq:ExpVar}}
   \label{tab:Dcoeff}
   \end{center}
\end{table*}

\bibliographystyle{apsrev4-1}
\bibliography{my_bib}

\begin{thebibliography}{28}%
\makeatletter
\providecommand \@ifxundefined [1]{%
 \@ifx{#1\undefined}
}%
\providecommand \@ifnum [1]{%
 \ifnum #1\expandafter \@firstoftwo
 \else \expandafter \@secondoftwo
 \fi
}%
\providecommand \@ifx [1]{%
 \ifx #1\expandafter \@firstoftwo
 \else \expandafter \@secondoftwo
 \fi
}%
\providecommand \natexlab [1]{#1}%
\providecommand \enquote  [1]{``#1''}%
\providecommand \bibnamefont  [1]{#1}%
\providecommand \bibfnamefont [1]{#1}%
\providecommand \citenamefont [1]{#1}%
\providecommand \href@noop [0]{\@secondoftwo}%
\providecommand \href [0]{\begingroup \@sanitize@url \@href}%
\providecommand \@href[1]{\@@startlink{#1}\@@href}%
\providecommand \@@href[1]{\endgroup#1\@@endlink}%
\providecommand \@sanitize@url [0]{\catcode `\\12\catcode `\$12\catcode
  `\&12\catcode `\#12\catcode `\^12\catcode `\_12\catcode `\%12\relax}%
\providecommand \@@startlink[1]{}%
\providecommand \@@endlink[0]{}%
\providecommand \url  [0]{\begingroup\@sanitize@url \@url }%
\providecommand \@url [1]{\endgroup\@href {#1}{\urlprefix }}%
\providecommand \urlprefix  [0]{URL }%
\providecommand \Eprint [0]{\href }%
\providecommand \doibase [0]{http://dx.doi.org/}%
\providecommand \selectlanguage [0]{\@gobble}%
\providecommand \bibinfo  [0]{\@secondoftwo}%
\providecommand \bibfield  [0]{\@secondoftwo}%
\providecommand \translation [1]{[#1]}%
\providecommand \BibitemOpen [0]{}%
\providecommand \bibitemStop [0]{}%
\providecommand \bibitemNoStop [0]{.\EOS\space}%
\providecommand \EOS [0]{\spacefactor3000\relax}%
\providecommand \BibitemShut  [1]{\csname bibitem#1\endcsname}%
\let\auto@bib@innerbib\@empty
%</preamble>
\bibitem [{\citenamefont {DeWitt}(1975)}]{dewitt1975quantum}%
  \BibitemOpen
  \bibfield  {author} {\bibinfo {author} {\bibfnamefont {B.~S.}\ \bibnamefont
  {DeWitt}},\ }\href@noop {} {\bibfield  {journal} {\bibinfo  {journal}
  {Physics Reports}\ }\textbf {\bibinfo {volume} {19}},\ \bibinfo {pages} {295}
  (\bibinfo {year} {1975})}\BibitemShut {NoStop}%
\bibitem [{\citenamefont {Christensen}(1976)}]{ChristensenPointSplit}%
  \BibitemOpen
  \bibfield  {author} {\bibinfo {author} {\bibfnamefont {S.~M.}\ \bibnamefont
  {Christensen}},\ }\href {\doibase 10.1103/PhysRevD.14.2490} {\bibfield
  {journal} {\bibinfo  {journal} {Phys. Rev. D}\ }\textbf {\bibinfo {volume}
  {14}},\ \bibinfo {pages} {2490} (\bibinfo {year} {1976})}\BibitemShut
  {NoStop}%
\bibitem [{\citenamefont {Candelas}\ and\ \citenamefont
  {Howard}(1984)}]{CandelasHowardPhi2}%
  \BibitemOpen
  \bibfield  {author} {\bibinfo {author} {\bibfnamefont {P.}~\bibnamefont
  {Candelas}}\ and\ \bibinfo {author} {\bibfnamefont {K.~W.}\ \bibnamefont
  {Howard}},\ }\href {\doibase 10.1103/PhysRevD.29.1618} {\bibfield  {journal}
  {\bibinfo  {journal} {Phys. Rev. D}\ }\textbf {\bibinfo {volume} {29}},\
  \bibinfo {pages} {1618} (\bibinfo {year} {1984})}\BibitemShut {NoStop}%
\bibitem [{\citenamefont {Ottewill}\ and\ \citenamefont
  {Taylor}(2010)}]{OttewillTaylorCS}%
  \BibitemOpen
  \bibfield  {author} {\bibinfo {author} {\bibfnamefont {A.~C.}\ \bibnamefont
  {Ottewill}}\ and\ \bibinfo {author} {\bibfnamefont {P.}~\bibnamefont
  {Taylor}},\ }\href {\doibase 10.1103/PhysRevD.82.104013} {\bibfield
  {journal} {\bibinfo  {journal} {Phys. Rev. D}\ }\textbf {\bibinfo {volume}
  {82}},\ \bibinfo {pages} {104013} (\bibinfo {year} {2010})}\BibitemShut
  {NoStop}%
\bibitem [{\citenamefont {Levi}\ and\ \citenamefont {Ori}(2015)}]{LeviOriMSP}%
  \BibitemOpen
  \bibfield  {author} {\bibinfo {author} {\bibfnamefont {A.}~\bibnamefont
  {Levi}}\ and\ \bibinfo {author} {\bibfnamefont {A.}~\bibnamefont {Ori}},\
  }\href {\doibase 10.1103/PhysRevD.91.104028} {\bibfield  {journal} {\bibinfo
  {journal} {Phys. Rev. D}\ }\textbf {\bibinfo {volume} {91}},\ \bibinfo
  {pages} {104028} (\bibinfo {year} {2015})}\BibitemShut {NoStop}%
\bibitem [{\citenamefont {Taylor}\ and\ \citenamefont
  {Breen}(2016)}]{TaylorBreenEvenD}%
  \BibitemOpen
  \bibfield  {author} {\bibinfo {author} {\bibfnamefont {P.}~\bibnamefont
  {Taylor}}\ and\ \bibinfo {author} {\bibfnamefont {C.}~\bibnamefont {Breen}},\
  }\href@noop {} {\  (\bibinfo {year} {2016})},\ \Eprint
  {http://arxiv.org/abs/In preparation} {In preparation} \BibitemShut {NoStop}%
\bibitem [{\citenamefont {Candelas}(1980)}]{Candelas:1980zt}%
  \BibitemOpen
  \bibfield  {author} {\bibinfo {author} {\bibfnamefont {P.}~\bibnamefont
  {Candelas}},\ }\href {\doibase 10.1103/PhysRevD.21.2185} {\bibfield
  {journal} {\bibinfo  {journal} {Phys. Rev. D}\ }\textbf {\bibinfo {volume}
  {21}},\ \bibinfo {pages} {2185} (\bibinfo {year} {1980})}\BibitemShut
  {NoStop}%
\bibitem [{\citenamefont {Candelas}\ and\ \citenamefont
  {Jensen}(1986)}]{Candelas:1985ip}%
  \BibitemOpen
  \bibfield  {author} {\bibinfo {author} {\bibfnamefont {P.}~\bibnamefont
  {Candelas}}\ and\ \bibinfo {author} {\bibfnamefont {B.~P.}\ \bibnamefont
  {Jensen}},\ }\href {\doibase 10.1103/PhysRevD.33.1596} {\bibfield  {journal}
  {\bibinfo  {journal} {Phys. Rev. D}\ }\textbf {\bibinfo {volume} {33}},\
  \bibinfo {pages} {1596} (\bibinfo {year} {1986})}\BibitemShut {NoStop}%
\bibitem [{\citenamefont {Anderson}(1989)}]{Anderson:1989vg}%
  \BibitemOpen
  \bibfield  {author} {\bibinfo {author} {\bibfnamefont {P.~R.}\ \bibnamefont
  {Anderson}},\ }\href {\doibase 10.1103/PhysRevD.39.3785} {\bibfield
  {journal} {\bibinfo  {journal} {Phys. Rev. D}\ }\textbf {\bibinfo {volume}
  {39}},\ \bibinfo {pages} {3785} (\bibinfo {year} {1989})}\BibitemShut
  {NoStop}%
\bibitem [{\citenamefont {DeBenedictis}(1999)}]{DeBenedictis:1998be}%
  \BibitemOpen
  \bibfield  {author} {\bibinfo {author} {\bibfnamefont {A.}~\bibnamefont
  {DeBenedictis}},\ }\href {\doibase 10.1023/A:1026734521256} {\bibfield
  {journal} {\bibinfo  {journal} {Gen.~Rel.~Grav.}\ }\textbf {\bibinfo {volume}
  {31}},\ \bibinfo {pages} {1549} (\bibinfo {year} {1999})},\ \Eprint
  {http://arxiv.org/abs/gr-qc/9804032} {arXiv:gr-qc/9804032 [gr-qc]}
  \BibitemShut {NoStop}%
%%CITATION = GR-QC/9804032;%%
\bibitem [{\citenamefont {Frolov}(1982)}]{Frolov:1982pi}%
  \BibitemOpen
  \bibfield  {author} {\bibinfo {author} {\bibfnamefont {V.~P.}\ \bibnamefont
  {Frolov}},\ }\href {\doibase 10.1103/PhysRevD.26.954} {\bibfield  {journal}
  {\bibinfo  {journal} {Phys.~Rev.}\ }\textbf {\bibinfo {volume} {D 26}},\
  \bibinfo {pages} {954} (\bibinfo {year} {1982})}\BibitemShut {NoStop}%
%%CITATION = PHRVA,D26,954;%%
\bibitem [{\citenamefont {Breen}\ and\ \citenamefont
  {Ottewill}(2010)}]{breen:10}%
  \BibitemOpen
  \bibfield  {author} {\bibinfo {author} {\bibfnamefont {C.}~\bibnamefont
  {Breen}}\ and\ \bibinfo {author} {\bibfnamefont {A.~C.}\ \bibnamefont
  {Ottewill}},\ }\href {\doibase 10.1103/PhysRevD.82.084019} {\bibfield
  {journal} {\bibinfo  {journal} {Phys. Rev. D}\ }\textbf {\bibinfo {volume}
  {82}},\ \bibinfo {pages} {084019} (\bibinfo {year} {2010})}\BibitemShut
  {NoStop}%
\bibitem [{\citenamefont {Winstanley}\ and\ \citenamefont
  {Young}(2008)}]{Winstanley:2007tf}%
  \BibitemOpen
  \bibfield  {author} {\bibinfo {author} {\bibfnamefont {E.}~\bibnamefont
  {Winstanley}}\ and\ \bibinfo {author} {\bibfnamefont {P.~M.}\ \bibnamefont
  {Young}},\ }\href {\doibase 10.1103/PhysRevD.77.024008} {\bibfield  {journal}
  {\bibinfo  {journal} {Phys.~Rev.}\ }\textbf {\bibinfo {volume} {D 77}},\
  \bibinfo {pages} {024008} (\bibinfo {year} {2008})},\ \Eprint
  {http://arxiv.org/abs/0708.3820} {arXiv:0708.3820 [gr-qc]} \BibitemShut
  {NoStop}%
%%CITATION = ARXIV:0708.3820;%%
\bibitem [{\citenamefont {Flachi}\ and\ \citenamefont
  {Tanaka}(2008)}]{Flachi:2008sr}%
  \BibitemOpen
  \bibfield  {author} {\bibinfo {author} {\bibfnamefont {A.}~\bibnamefont
  {Flachi}}\ and\ \bibinfo {author} {\bibfnamefont {T.}~\bibnamefont
  {Tanaka}},\ }\href {\doibase 10.1103/PhysRevD.78.064011} {\bibfield
  {journal} {\bibinfo  {journal} {Phys.~Rev.}\ }\textbf {\bibinfo {volume} {D
  78}},\ \bibinfo {pages} {064011} (\bibinfo {year} {2008})},\ \Eprint
  {http://arxiv.org/abs/0803.3125} {arXiv:0803.3125 [hep-th]} \BibitemShut
  {NoStop}%
%%CITATION = ARXIV:0803.3125;%%
\bibitem [{\citenamefont {Morgan}\ \emph {et~al.}(2007)\citenamefont {Morgan},
  \citenamefont {Thom}, \citenamefont {Winstanley},\ and\ \citenamefont
  {Young}}]{Morgan:2007hp}%
  \BibitemOpen
  \bibfield  {author} {\bibinfo {author} {\bibfnamefont {D.}~\bibnamefont
  {Morgan}}, \bibinfo {author} {\bibfnamefont {S.}~\bibnamefont {Thom}},
  \bibinfo {author} {\bibfnamefont {E.}~\bibnamefont {Winstanley}}, \ and\
  \bibinfo {author} {\bibfnamefont {P.~M.}\ \bibnamefont {Young}},\ }\href
  {\doibase 10.1007/s10714-007-0486-3} {\bibfield  {journal} {\bibinfo
  {journal} {Gen.~Rel.~Grav.}\ }\textbf {\bibinfo {volume} {39}},\ \bibinfo
  {pages} {1719} (\bibinfo {year} {2007})},\ \Eprint
  {http://arxiv.org/abs/0705.1131} {arXiv:0705.1131 [gr-qc]} \BibitemShut
  {NoStop}%
%%CITATION = ARXIV:0705.1131;%%
\bibitem [{\citenamefont {D\'ecanini}\ and\ \citenamefont
  {Folacci}(2008)}]{DecaniniFolacciHadamardRen}%
  \BibitemOpen
  \bibfield  {author} {\bibinfo {author} {\bibfnamefont {Y.}~\bibnamefont
  {D\'ecanini}}\ and\ \bibinfo {author} {\bibfnamefont {A.}~\bibnamefont
  {Folacci}},\ }\href {\doibase 10.1103/PhysRevD.78.044025} {\bibfield
  {journal} {\bibinfo  {journal} {Phys. Rev. D}\ }\textbf {\bibinfo {volume}
  {78}},\ \bibinfo {pages} {044025} (\bibinfo {year} {2008})}\BibitemShut
  {NoStop}%
\bibitem [{\citenamefont {Thompson}\ and\ \citenamefont
  {Lemos}(2009)}]{Thompson:2008bk}%
  \BibitemOpen
  \bibfield  {author} {\bibinfo {author} {\bibfnamefont {R.~T.}\ \bibnamefont
  {Thompson}}\ and\ \bibinfo {author} {\bibfnamefont {J.~P.~S.}\ \bibnamefont
  {Lemos}},\ }\href {\doibase 10.1103/PhysRevD.80.064017} {\bibfield  {journal}
  {\bibinfo  {journal} {Phys.~Rev.}\ }\textbf {\bibinfo {volume} {D 80}},\
  \bibinfo {pages} {064017} (\bibinfo {year} {2009})},\ \Eprint
  {http://arxiv.org/abs/0811.3962} {arXiv:0811.3962 [gr-qc]} \BibitemShut
  {NoStop}%
%%CITATION = ARXIV:0811.3962;%%
\bibitem [{\citenamefont {Matyjasek}\ and\ \citenamefont
  {Sadurski}(2015{\natexlab{a}})}]{MatyjasekHD1}%
  \BibitemOpen
  \bibfield  {author} {\bibinfo {author} {\bibfnamefont {J.}~\bibnamefont
  {Matyjasek}}\ and\ \bibinfo {author} {\bibfnamefont {P.}~\bibnamefont
  {Sadurski}},\ }\href {\doibase 10.1103/PhysRevD.92.044023} {\bibfield
  {journal} {\bibinfo  {journal} {Phys. Rev. D}\ }\textbf {\bibinfo {volume}
  {92}},\ \bibinfo {pages} {044023} (\bibinfo {year}
  {2015}{\natexlab{a}})}\BibitemShut {NoStop}%
\bibitem [{\citenamefont {Matyjasek}\ and\ \citenamefont
  {Sadurski}(2015{\natexlab{b}})}]{MatyjasekHD2}%
  \BibitemOpen
  \bibfield  {author} {\bibinfo {author} {\bibfnamefont {J.}~\bibnamefont
  {Matyjasek}}\ and\ \bibinfo {author} {\bibfnamefont {P.}~\bibnamefont
  {Sadurski}},\ }\href {\doibase 10.1103/PhysRevD.91.044027} {\bibfield
  {journal} {\bibinfo  {journal} {Phys. Rev. D}\ }\textbf {\bibinfo {volume}
  {91}},\ \bibinfo {pages} {044027} (\bibinfo {year}
  {2015}{\natexlab{b}})}\BibitemShut {NoStop}%
\bibitem [{\citenamefont {{Frolov}}\ \emph {et~al.}(1989)\citenamefont
  {{Frolov}}, \citenamefont {{Mazzitelli}},\ and\ \citenamefont
  {{Paz}}}]{Fro:89}%
  \BibitemOpen
  \bibfield  {author} {\bibinfo {author} {\bibfnamefont {V.~P.}\ \bibnamefont
  {{Frolov}}}, \bibinfo {author} {\bibfnamefont {F.~D.}\ \bibnamefont
  {{Mazzitelli}}}, \ and\ \bibinfo {author} {\bibfnamefont {J.~P.}\
  \bibnamefont {{Paz}}},\ }\href {\doibase 10.1103/PhysRevD.40.948} {\bibfield
  {journal} {\bibinfo  {journal} {\prd}\ }\textbf {\bibinfo {volume} {40}},\
  \bibinfo {pages} {948} (\bibinfo {year} {1989})}\BibitemShut {NoStop}%
\bibitem [{\citenamefont {Flachi}\ \emph {et~al.}(2016)\citenamefont {Flachi},
  \citenamefont {Quinta},\ and\ \citenamefont {Lemos}}]{Flachi:HD2016}%
  \BibitemOpen
  \bibfield  {author} {\bibinfo {author} {\bibfnamefont {A.}~\bibnamefont
  {Flachi}}, \bibinfo {author} {\bibfnamefont {G.~M.}\ \bibnamefont {Quinta}},
  \ and\ \bibinfo {author} {\bibfnamefont {J.~P.~S.}\ \bibnamefont {Lemos}},\
  }\href@noop {} {\  (\bibinfo {year} {2016})},\ \Eprint
  {http://arxiv.org/abs/arXiv:1609.06794} {arXiv:1609.06794} \BibitemShut
  {NoStop}%
\bibitem [{\citenamefont {Breen}\ \emph {et~al.}(2015)\citenamefont {Breen},
  \citenamefont {Hewitt}, \citenamefont {Winstanley},\ and\ \citenamefont
  {Ottewill}}]{breen2015vacuum}%
  \BibitemOpen
  \bibfield  {author} {\bibinfo {author} {\bibfnamefont {C.}~\bibnamefont
  {Breen}}, \bibinfo {author} {\bibfnamefont {M.}~\bibnamefont {Hewitt}},
  \bibinfo {author} {\bibfnamefont {E.}~\bibnamefont {Winstanley}}, \ and\
  \bibinfo {author} {\bibfnamefont {A.~C.}\ \bibnamefont {Ottewill}},\ }\href
  {\doibase 10.1103/PhysRevD.92.084039} {\bibfield  {journal} {\bibinfo
  {journal} {Phys. Rev. D}\ }\textbf {\bibinfo {volume} {92}},\ \bibinfo
  {pages} {084039} (\bibinfo {year} {2015})}\BibitemShut {NoStop}%
\bibitem [{\citenamefont {Shiraishi}\ and\ \citenamefont
  {Maki}(1994)}]{shiraishi1994vacuum}%
  \BibitemOpen
  \bibfield  {author} {\bibinfo {author} {\bibfnamefont {K.}~\bibnamefont
  {Shiraishi}}\ and\ \bibinfo {author} {\bibfnamefont {T.}~\bibnamefont
  {Maki}},\ }\href {http://stacks.iop.org/0264-9381/11/i=7/a=009} {\bibfield
  {journal} {\bibinfo  {journal} {Classical and Quantum Gravity}\ }\textbf
  {\bibinfo {volume} {11}},\ \bibinfo {pages} {1687} (\bibinfo {year}
  {1994})}\BibitemShut {NoStop}%
\bibitem [{\citenamefont {Wald}(1994)}]{WaldQFT}%
  \BibitemOpen
  \bibfield  {author} {\bibinfo {author} {\bibfnamefont {R.~M.}\ \bibnamefont
  {Wald}},\ }\href@noop {} {\emph {\bibinfo {title} {Quantum Field Theory in
  Curved Spacetime and Black Hole Thermodynamics}}}\ (\bibinfo  {publisher}
  {University of Chicago Press},\ \bibinfo {year} {1994})\BibitemShut {NoStop}%
\bibitem [{MyW()}]{MyWebpage}%
  \BibitemOpen
  \href@noop {} {}\bibinfo {howpublished}
  {\url{http://www.taylorexpansion.net}}\BibitemShut {NoStop}%
\bibitem [{\citenamefont {Cohl}(2013)}]{CohlGegenInt}%
  \BibitemOpen
  \bibfield  {author} {\bibinfo {author} {\bibfnamefont {H.~S.}\ \bibnamefont
  {Cohl}},\ }\href {\doibase 10.1080/10652469.2012.761613} {\bibfield
  {journal} {\bibinfo  {journal} {Integral Transforms and Special Functions}\
  }\textbf {\bibinfo {volume} {24}},\ \bibinfo {pages} {807} (\bibinfo {year}
  {2013})}\BibitemShut {NoStop}%
\bibitem [{\citenamefont {Gradshteyn}\ and\ \citenamefont
  {Ryzhik}(2000)}]{GradRiz}%
  \BibitemOpen
  \bibfield  {author} {\bibinfo {author} {\bibfnamefont {I.~S.}\ \bibnamefont
  {Gradshteyn}}\ and\ \bibinfo {author} {\bibfnamefont {I.~M.}\ \bibnamefont
  {Ryzhik}},\ }\href@noop {} {\emph {\bibinfo {title} {Table of Integrals,
  Series and Products}}}\ (\bibinfo  {publisher} {Academic Press},\ \bibinfo
  {year} {2000})\BibitemShut {NoStop}%
\bibitem [{\citenamefont {Hewitt}(2015)}]{Hewitt:15}%
  \BibitemOpen
  \bibfield  {author} {\bibinfo {author} {\bibfnamefont {M.}~\bibnamefont
  {Hewitt}},\ }\href@noop {} {Ph.D. thesis},\ \bibinfo  {school} {{U}niversity
  of {S}heffield} (\bibinfo {year} {2015})\BibitemShut {NoStop}%
\end{thebibliography}%

\end{document}